\documentclass{emulateapj}
\usepackage{amsmath, natbib, ulem}
\usepackage[usenames,dvips]{color}

\slugcomment{Submitted to The Astrophysical Journal}

\begin{document}

\title{Angular Momentum Exchange in White Dwarf Binaries Accreting 
Through Direct Impact}

\author{J. F. Sepinsky$^{1}$ and V. Kalogera$^2$}
\affil{$^1$ The University of Scranton, Department of Physics and
Electrical Engineering, Scranton, PA 18510, {\it 
jeremy.sepinsky@scranton.edu},}
\affil{$^2$ Center for Interdisciplinary Exploration and Research in 
Astrophysics (CIERA) \&  Department of Physics and Astronomy, Northwestern
University, 2145 Sheridan Road, Evanston, IL 60208, {\it 
vicky@northwestern.edu}}
\shorttitle{Angular Momentum Exchange}

\begin{abstract}

We examine the exchange of angular momentum between the component spins 
and the orbit in semi-detached double white dwarf binaries undergoing 
mass transfer through direct impact of the transfer stream.  We 
approximate the stream as a series of discrete massive particles ejected 
in the ballistic limit at the inner Lagrangian point of the donor toward 
the accretor.  This work improves upon similar earlier studies in a 
number of ways.  First, we self-consistently calculate the total angular 
momentum of the orbit at all times.  This includes changes in the 
orbital angular momentum during the ballistic trajectory of the ejected 
mass, as well as changes during the ejection/accretion due to the radial 
component of the particle's velocity.  Second, we calculate the 
particle's ballistic trajectory for each system, which allows us to 
determine the precise position and velocity of the particle upon 
accretion.  We can then include specific information about the radius of 
the accretor as well as the angle of impact.  Finally, we ensure that 
the total angular momentum is conserved, which requires the donor star 
spin to vary self-consistently. With these improvements we calculate the 
angular momentum change of the orbit and each binary component across 
the entire parameter space of direct impact double white dwarf binary 
systems.  We find a significant decrease in the amount of angular 
momentum removed from the orbit during mass transfer, as well as cases 
where this process increases the angular momentum of the orbit at the 
expense of the spin angular momentum of the donor.  We conclude that, 
unlike earlier claims in the literature, mass transfer through direct 
impact need not destabilize the binary and that the quantity and sign of 
the orbital angular momentum transfer depends on the binary properties, 
particularly the masses of the double white dwarf binary component 
stars.  This stabilization may significantly impact the population 
synthesis calculations of the expected numbers of events/systems for 
which double white dwarfs may be a progenitor, e.g. Type Ia supernovae, 
Type .Ia supernovae, and AM CVn.

\end{abstract}


\section{Introduction}
\label{sec-intro}

Double white dwarf (DWD) binaries provide an interesting key to 
understanding a number of different astrophysical problems.  Their birth 
properties provide insight into the evolution of their progenitors 
\citep[e.g., ][]{2001A&A...375..890N}, as well as the dynamics of common 
envelope evolution \citep[e.g., ][]{2012A&A...546A..70T}.  From a 
population standpoint, DWDs may make up a large fraction of the close 
binary stars in our galaxy, creating a confusion limited background for 
high frequency space-based gravitational wave detectors such as LISA 
\citep{2001A&A...375..890N, 2010ApJ...717.1006R}.  As these objects 
evolve via gravitational radiation loses \citep{1984ApJ...277..355W, 
1984ApJS...54..335I} their shrinking orbit may eventually allow them to 
be detected as individual systems \citep[e.g., ][]{1981ApJ...244..269N, 
1987ApJ...323..129E, 1990ApJ...360...75H, 1988ApJ...334..947S}.

As the DWD orbit continues to shrink via gravitational radiation, the 
less massive component will inevitably fill its Roche lobe and the 
system will enter a semi-detached state.  It is in this interacting 
phase when sources and sinks of angular momentum other than 
gravitational radiation become vitally important in determining the 
eventual fate of the system.  \citet{2004MNRAS.350..113M} \citep[See 
also][in prep]{2007ApJ...655.1010G, 1999A&A...349L..17H, Kremer2013} 
analyze the stability of the mass transfer process for DWDs.  A number 
of possibilities have been discussed in the literature.  If the mass 
transfer process if unstable, the binary orbit will shrink until a 
merger occurs.  Such a merger may create such objects as R CrB or SdB 
stars \citep[e.g., ][]{2002MNRAS.333..121S}.  If the system is 
sufficiently massive, a Type Ia supernova may result 
\citep{1984ApJ...277..355W, 1984ApJS...54..335I, 2010ApJ...722L.157V, 
2012A&A...546A..70T}.  Mergers of low-mass white dwarfs have also been 
proposed as the origin of Type .Ia supernovae 
\citep{2010ApJ...715..767S, 2013arXiv1310.6359K}, where the properties 
of the binary immediately prior to merger can significantly affect the 
observed characteristics of such explosion \citep{2010ApJ...722L.157V}. 
In the case where the mass transfer is stable, such systems may be 
identifiable as AM CVn \citep{1996MNRAS.280.1035T, 2001A&A...368..939N}.

Three such stable double white dwarf AM CVn are thought to exist: HM 
Cnc, V407 Vul, and EX Cet \citep[See][and references 
therein]{2010PASP..122.1133S}.  Each of these systems has an orbital 
period less than 12 minutes with no observable disk, making them 
excellent candidates for direct impact (DI) mass transfer.  
\citet{2001A&A...368..939N} noted that, since the radius of the accretor 
in a DWD is large compared to the orbital separation, DI mass 
transfer is likely to occur.  Indeed, examining the distance of closest 
approach for the ballistic trajectories of \citet{1975ApJ...198..383L}, 
inserting the Roche lobe radius of \citet{1983ApJ...268..368E} and the 
mass-radius relation for zero-temperature white dwarfs of 
\citet{1988ApJ...332..193V}, one finds \citep[see Figure~1 
of][Figure~\ref{fig-di} of this paper]{2004MNRAS.350..113M} that the 
majority of the parameter space of DWDs will undergo DI mass transfer.

The evolution of the binary through this DI mass transfer phase 
crucially depends on the exchange of angular momentum between the 
binary components, as well as between the orbit and the component spins.  
Understanding this process is necessary to determine accurate rates for 
the creation of many of the astrophysical phenomena discussed 
previously.

\citet{2004MNRAS.350..113M} finds that, while the stability of the 
system is largely dependent on the unknown tidal coupling between the 
accretor and the orbit, the majority of DI systems will be 
unstable.  \citet{2007ApJ...655.1010G} notes that the stability of these 
systems may be improved if the rotation rate of the donor is allowed to 
vary.  However, both of these studies assume that (i) the orbital 
angular momentum of the particle at accretion is equal to the angular 
momentum of the circularization radius \citep{1988ApJ...332..193V, 
1975MNRAS.170..325F}, (ii) that all the angular momentum of the particle 
is deposited enitrely into the spin of the accretor, and (iii) that the 
angular momentum change during the particle's motion can be extracted 
from the semi-major axis of the binary orbit.  As we show here, these 
assumptions are not always justified.  Indeed, hydrodynamic studies of 
close white dwarf binaries suggest that these systems my be more stable 
than previously expected \citep{2007ApJ...670.1314M, 
2009JPhCS.172a2034D}.

In this paper, we follow the full ballistic trajectory of the ejected 
mass for each system, calculating the feedback onto the spin and orbital 
angular momenta of each star throughout the ejection, transit, and 
accretion process.  In section~\ref{sec-ballist}, we describe our 
ballistic model and highlight the changes in the angular momenta of the 
donor and accretor.  In section~\ref{sec-angmom} we examine the total 
changes in the angular momenta of the system for any DWD undergoing 
DI mass transfer and compare to previous results.  Finally, 
in section~\ref{sec-disc}, we discuss the implications of these findings 
and propose a method to employ these results in future studies.

\section{Ballistic Trajectories}
\label{sec-ballist}

In this section, we describe the ballistic models of 
\citet{2010ApJ...724..546S} as used here for the particular case of DWD 
Roche lobe overflow DI mass transfer.  Below, we highlight the portions 
of the calculation that are particularly important to this problem and 
refer the interested reader to that paper and references therein for a 
more general discussion of orbital evolution due to DI mass transfer.

\subsection{Basic Assumptions}

We consider a close binary system of two white dwarfs with masses $M_D$ 
(donor) and $M_A$ (accretor), volume-equivalent radii of ${\cal R}_D$ 
and ${\cal R}_A$, and uniform rotation rates $\vec{\Omega}_D$ and 
$\vec{\Omega}_A$ with axes perpendicular to the orbital plane.  We 
assume the mass of each star is distributed spherically symmetrically.  
The binary is assumed to be in an initially circular Keplerian orbit 
with semi-major axis $a$ and orbital period $P_{\rm orb}$.
The radius of each object 
is assigned following Eggleton's zero-temperature mass-radius relation 
\citep[equation~15 of][]{1988ApJ...332..193V}.  We chose the 
semi-major axis of the orbit such that the volume equivalent radius of 
the mass donor (${\cal R}_D$) is equal to the volume-equivalent radius 
of its Roche lobe as fit by \citet{1983ApJ...268..368E}.  Both the donor 
and accretor initially rotate synchronously with the orbit\footnote{If 
the donor rotates non-synchronously, the Roche lobe radii can be 
calculated as given in \citet{2007ApJ...660.1624S}.}.

Our ballistic calculations are performed in a stationary inertial 
reference frame located at the initial center of mass of the binary 
system.  From this reference frame, the positions of the centers of mass 
the donor and accretor are given by $\vec{R}_D$ and $\vec{R}_A$ with 
velocities $\vec{V}_D$ and $\vec{V}_A$, respectively.

\subsection{Ballistic Model}

We model the mass transfer as a discrete event where a particle of mass 
$M_P \ll M_D,M_A$ is instantaneously ejected from the inner Lagrangian 
point of the donor, $\vec{r}_{L_1}$ (measured from the center of mass 
of the donor), with velocity $\vec{V}_P$ equal to the vector sum of the 
donor's rotational velocity at the point of ejection, the orbital 
velocity of the donor, and the ejection velocity of the particle 
relative to the center of mass of star~1, $\vec{V}_{\rm ej}$,
\begin{equation}
\vec{V}_P = \vec{\Omega}_D\times\vec{r}_{L_1}+\vec{V}_D + 
\vec{V}_{\rm 
 ej}.
\end{equation}
We let $\vec{V}_{\rm ej}$ have a magnitude equal to the sound speed in a 
white dwarf atmosphere of temperature $20\ 000\,K$ directed along the 
line connecting the mass centers of the two objects.  We note that, 
provided $|\vec{V}_{\rm ej}|/|\vec{V}_P| \lesssim 0.01$ the angular 
momentum exchange between the particle and the system during transport 
is unaffected by changes in $|\vec{V}_{\rm ej}|$.

When the particle is ejected, the center of mass of the system remains 
fixed.  Since the particle's mass is instantaneously transported to the 
surface, this forces a shift in the position of the donor's center of 
mass.  Additionally, conservation of linear and angular momentum imply a 
change in the donor's velocity and rotation rate, respectively.  Writing 
these conservation laws yields the following equations for the position, 
velocity, and rotation rate of the donor after ejection
\begin{eqnarray}
 \vec{R}_{D,F}&=&M_{D,F}^{-1}\left(M_{D,I}\vec{R}_{D,I}-M_P\vec{R}_P\right)
  \label{eq-o1}  \\
 \vec{V}_{D,F}&=&M_{D,F}^{-1}\left( 
 M_{D,I}\vec{V}_{D,I}-M_P\vec{V}_P\right) \\
 \vec{\Omega}_{D,F} &=& I_{D,F}^{-1}\left( I_{D,I}\vec{\Omega}_{D,I} + 
 M_{D,I}\vec{R}_{D,I}\times\vec{V}_{D,I}\right. \nonumber\\
 &&\left. -M_{D,F}\vec{R}_{D,F}\times\vec{V}_{D,F} 
 -M_P\vec{R}_P\times\vec{V}_P \right) \label{eq-o3}
\end{eqnarray}
where $I_D$ is the moment of inertia of the donor star.  The subscript 
``I" denotes that the quantity is considered immediately before the 
ejection of the particle, and the subscript ``F'' denotes the quantity 
immediately after ejection.

Following ejection, the trajectory of the three bodies are numerically 
integrated following the standard Newtonian equations of motion using an 
$8^{\rm th}$ order Runge Kutta ordinary differential equation solver 
\citep{GNUGSL}.  Throughout the course of the integration, the total 
energy and momentum of the system are conserved to, generally, better 
than 1 part in $10^{12}$.  While there is no coupling between the spins 
of either component and the motion of the ejected particle, the orbital 
angular momentum of both donor and accretor will be affected by the 
transport of the particle.

During the particle's trajectory, if it ever comes closer to the 
accretor than its radius, i.e., $|\vec{R}_P-\vec{R}_A| < {\cal R}_A$, 
the particle accretes.  When this occurs, the center of mass of the 
system remains fixed.  Since the particle's mass is added to the 
accretor, this necessitates a shift in the accretor's center of mass.  
Additionally, conservation of linear and angular momentum imply a change 
in the accretor's velocity and rotation rate, respectively.  To 
calculate the change in these quantities during accretion it is 
sufficient to use equations~(\ref{eq-o1})--(\ref{eq-o3}), simply 
replacing all quantities referencing the donor with the respective 
quantities referencing the accretor and letting $M_P \rightarrow -M_P$.  
In this case, the position and velocity of the particle are those 
immediately prior to accretion.

\subsection{Binary Parameter Space for Mass Transfer through Direct 
Impact}
\label{sec-di}

\begin{figure}
\plotone{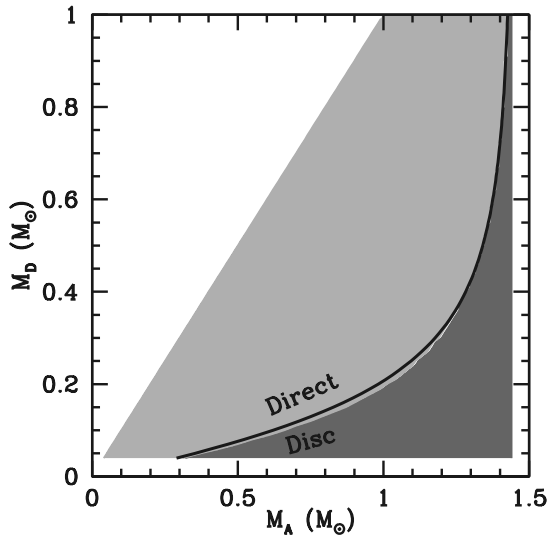}
\caption{The result of a single ballistic integration as a function of 
the masses of the donor and accretor.  Light grey points represent 
systems which undergo DI, while dark grey points undergo DF.  The solid 
line represents the analytical boundary as described in 
\citet{2004MNRAS.350..113M} (See text for details).}
\label{fig-di}
\end{figure}

For the purpose of this work, we are only interested in orbital changes 
due to ballistic trajectories where the particle impacts the surface of 
the accretor within one orbital period (i.e., those in which DI occurs).  
Following the above procedure, we compute the outcome for systems with 
$M_A < 1.44\,M_\odot$ and $M_D < M_A$.  If the particle accretes within 
one orbital period, we classify this as a direct impact.  If the 
particle does not accrete within one orbital period it is likely the 
ejection stream from the donor will self-intersect, resulting in the 
eventual formation of an accretion disc.  We identify these systems as 
disc formation (DF) systems\footnote{We note that in 
\citet{2010ApJ...724..546S} we identified a third possible outcome of 
this ejection scenario: Self-accretion, in which the ejected mass 
falls back onto the donor star.  This outcome is only possible for 
eccentric and/or non-synchronously rotating systems, both of which lie 
outside the parameter space of this study.}.

In Figure~\ref{fig-di} the light grey region corresponds to systems 
which undergo DI, while the dark grey region corresponds to systems 
which undergo DF.  The solid line near the boundary of these two regions 
is an analytical approximation to this boundary, which was first derived 
and presented by \citet{2004MNRAS.350..113M} (which itself followed a 
similar derivation in \citet{2001A&A...368..939N}).  Following them, we 
calculate this line by taking the distance of closest approach for a 
ballistic trajectory given by \citet{1975ApJ...198..383L} (as 
analytically fit by \citet{2001A&A...368..939N}) and setting it equal to 
the radius of the accretor (as given by Eggleton's zero-temperature 
mass-radius relation \citep{1988ApJ...332..193V}).  In calculating this 
line, we assume the donor (also described by the same zero-temperature 
mass-radius relation) completely fills its Roche lobe 
\citep{1983ApJ...268..368E}.  As expected, our numerical results agree 
quite well to this analytic approximation. We fit the transition between 
DI and DF as given by our numerical calculations with the equation
\begin{equation}
M_D = \frac{1-9.24M_A+4.75M_A^2}{36M_A-54}.
\label{eq-fit}
\end{equation}
This equation is plotted as the solid black line in 
Figure~\ref{fig-di-j}.

\subsection{Angular Momentum Changes Over One Orbit}
\label{sec-evo}

To calculate the stability of white dwarf binary systems previous 
studies \citep[e.g.,][]{2004MNRAS.350..113M, 2007ApJ...655.1010G} have 
focused primarily on changes in the orbital angular momentum of the 
system since such changes are directly associated with changes in the 
mass ratio and semi-major axis of the system.  In those studies, the 
angular momentum exchange was calculated using a numerical prescription 
based on \citet{1988ApJ...332..193V}; it assumed that the angular 
momentum transferred from the orbit to the spin of the accretor is 
exactly equal to the average angular momentum of the ballistic particle 
during its motion from the donor to the accretor.  Here we examine the 
angular momentum exchange between all relevant components: the two 
binary members and the orbit, and we find that for a certain part of the 
parameter space the usual assumption above does not hold. Specifically, 
we allow the spin of the donor to vary 
self-consistently\footnote{\citet{2007ApJ...655.1010G} does allow for 
variation in the spin of the donor, but does not do so in a a 
self-consistent manner.  They reduce the specific spin angular momentum 
of the donor by the specific angular momentum of the ejected mass on the 
surface of the star.}, and we account for the feedback onto the orbit at 
the moments of mass ejection and accretion.  We explain the derivation 
of the full account of angular-momentum exchange in what follows.

We begin by rewriting equations~(\ref{eq-o1})--(\ref{eq-o3}) in terms of 
the change in the position, velocity, and spin angular momentum of the 
donor, $\vec{J}_{{\rm Spin,} D}=I_D\vec{\Omega}_D$, per unit ejected 
mass.  To first order in the mass of the ballistic particle, we find
\begin{eqnarray}
\frac{\Delta \vec{R}_D}{M_P} &\equiv& 
\frac{\vec{R}_{D,F}-\vec{R}_{D,I}}{M_P} 
= \frac{\vec{R}_{D,I} - \vec{R}_P}{M_{D,I}} \label{eq-dr}\\
\frac{\Delta \vec{V}_D}{M_P} &\equiv& 
\frac{\vec{V}_{D,F}-\vec{V}_{D,I}}{M_P} 
= \frac{\vec{V}_{D,I}- \vec{V}_P}{M_{D,I}}\\
\frac{\Delta \vec{J}_{{\rm Spin,} D}}{M_P} &\equiv& \frac{\vec{J}_{{\rm 
Spin,} D, F} - \vec{J}_{{\rm Spin,} D, I}}{M_P} \nonumber \\ 
&=&-\left(\vec{R}_{D,I} - 
\vec{R}_{P}\right)\times\left(\vec{V}_{D,I}- \vec{V}_{P}\right). 
\label{eq-dj}
\end{eqnarray}

Since the total angular momentum of the system is conserved during 
ejection, it follows that
\begin{eqnarray}
 \Delta \vec{j}_{\rm ej}\equiv\frac{\Delta \vec{J}_{{\rm orb},D}}{M_P} 
 &\equiv& 
 \frac{\vec{J}_{{\rm 
 orb},D,F} - \vec{J}_{{\rm orb},D,I}}{M_P}\nonumber \\
 &=&-\frac{\Delta  \vec{J}_{{\rm  Spin,} D}}{M_P} - 
 \vec{R}_P\times\vec{V}_P
 \label{eq-ej}
\end{eqnarray}
where $\vec{R}_P\times\vec{V}_P$ is the specific angular momentum of the 
ejected particle as measured by the center of mass and we define $\Delta 
\vec{j}_{\rm ej}$ as the change in the orbital angular momentum per unit 
particle mass due to the ejection of a ballistic 
particle.\footnote{Alternatively, we can derive the final 
step of equation~\ref{eq-ej} as follows.  We first write the orbital 
angular momentum of the donor before and after accretion as 
$\vec{J}_{{\rm orb},D,F} = M_{D,F}\vec{R}_{D,F} \times \vec{V}_{D,F}$ 
and $\vec{J}_{{\rm orb},D,I} = M_{D,I}\vec{R}_{D,I} \times 
\vec{V}_{D,I}$, respectively. Then, by substituting 
equations~(\ref{eq-dr})--(\ref{eq-dj}) and recognizing that 
$M_{D,I}-M_{D,F} = M_P$, we can reproduce equation~(\ref{eq-ej}) without 
explicitly invoking the conservation of angular momentum.}

Similarly, during accretion
\begin{eqnarray}
 \Delta \vec{j}_{\rm ac}\equiv\frac{\Delta \vec{J}_{{\rm orb},A}}{M_P} 
 &\equiv& 
 \frac{\vec{J}_{{\rm 
 orb},A,F} - \vec{J}_{{\rm orb},A,I}}{M_P}\nonumber \\
 &=&-\frac{\Delta  \vec{J}_{{\rm  Spin,} A}}{M_P} + 
 \vec{R}_P^\prime\times\vec{V}_P^\prime
 \label{eq-ac}
\end{eqnarray}
where $\vec{R}_P^\prime\times\vec{V}_P^\prime$ is the specific angular 
momentum of the particle immediately before accretion as measured by the 
center of mass and we define $\Delta \vec{j}_{\rm ac}$ as the change in 
the orbital angular momentum per unit particle mass due to the accretion 
of a ballistic particle.  Equation~(\ref{eq-ac}) follows from rewriting 
equations~(\ref{eq-dr})-(\ref{eq-dj}) in terms of the accretor during 
particle impact by simply replacing all quantities referencing the donor 
with the respective quantities referencing the accretor and letting $M_P 
\rightarrow -M_P$.  In that derivation, the position and velocity of the 
particle are those immediately prior to accretion.

Since the total angular momentum must be conserved, the orbital angular 
momentum of the donor and accretor must vary as the orbital angular 
momentum of the particle varies during its transport from donor to 
accretor.  We do not include any coupling between the spins of the 
components and the particle trajectory (which could only arise due to 
tidal interactions between the ejected mass and the binary components), 
thus the orbital angular momentum is the only property affected.  We 
write the change of orbital angular momentum per unit particle mass 
during during transport from donor to accretor as $\Delta \vec{j}_{\rm 
t}$.

The total change in the orbital angular momentum per unit particle mass, 
$\Delta \vec{j}_{\rm orb, T}$, for a single ballistic trajectory is then
\begin{eqnarray}
 \Delta \vec{j}_{\rm orb, T} &=& \Delta \vec{j}_{\rm ej} + \Delta 
 \vec{j}_{\rm t} + \Delta \vec{j}_{\rm ac} \label {eq-jcon1}\\
 &=& -\Delta  \vec{j}_{{\rm  Spin,} D} -\Delta \vec{j}_{{\rm  Spin,} 
 A} \label{eq-jcon2},
\end{eqnarray}
where $\Delta \vec{j}_{{\rm  Spin,} D} = \Delta \vec{J}_{{\rm  Spin,} 
D}/M_P$ and $\Delta \vec{j}_{{\rm  Spin,} A} = \Delta \vec{J}_{{\rm  
Spin,} A}/M_P$ are the total change in the spin angular momentum of the 
donor and accretor, respectively, per unit particle mass.  This 
statement simply represents the conservation of angular momentum 
throughout all aspects of the ballistic calculation.

We note that neither \citet{2004MNRAS.350..113M} nor 
\citet{2007ApJ...655.1010G} include any immediate feedback on the orbit 
due to ejection and accretion.  As such, they impose $\Delta 
\vec{j}_{\rm ej} = \Delta \vec{j}_{\rm ac} = 0$.  This implies that all 
changes to the orbital angular momentum of the system must take place 
during the particle motion from donor to accretor.  Stated another way, 
equations~(\ref{eq-ej}) and (\ref{eq-ac}) show that this constraint 
implies that both ejection and accretion of the particle happen 
tangentially to the surface of the accretor, which is unlikely to be the 
case, particularly for the accretor \citep[See][]{1975ApJ...198..383L}.

For the component spins, \citet{2004MNRAS.350..113M} assume that the 
donor remains tidally locked, and thus $\Delta j_{{\rm Spin}, D}=0$, 
while \citet{2007ApJ...655.1010G} allow the spin of the donor to vary.  
Both papers assume that, during accretion, the orbital angular momentum 
of the particle is added entirely to the accretor's spin. But as 
equations~(\ref{eq-ej})-(\ref{eq-jcon1}) show, conservation of angular 
momentum then forces unrealistic constraints on the motion of the 
ballistic particle.  Specifically, the particle must follow a trajectory 
that starts with and results in tangential ejection and accretion, 
independent of the actual dynamics of the system, and must have 
velocities that exactly reproduce $\Delta j_{{\rm Spin}, D}$ and $\Delta 
j_{{\rm Spin}, A}$.  One should not expect such a trajectory to exist in 
the general case.  The change in the orbital angular momentum during 
particle motion is independent of the change in spins, and should be 
calculated separately.  We further enumerate the differences between our 
model and the previous results in section~\ref{sec-comp}.

In this paper, we directly calculate the change in orbital angular 
momentum at each stage -- ejection, particle motion, and accretion.  
While the orbital angular momentum of the particle need not be 
different in our two methods, the fraction of angular momentum that 
arises from, or is deposited into, the spin of the components may differ 
significantly.  In the next section, we explore the exchange of angular 
momentum for one example system in detail.

\subsection{Angular Momenta Variations Throughout One Orbit}
\label{sec-orbang}

\begin{figure}
\plotone{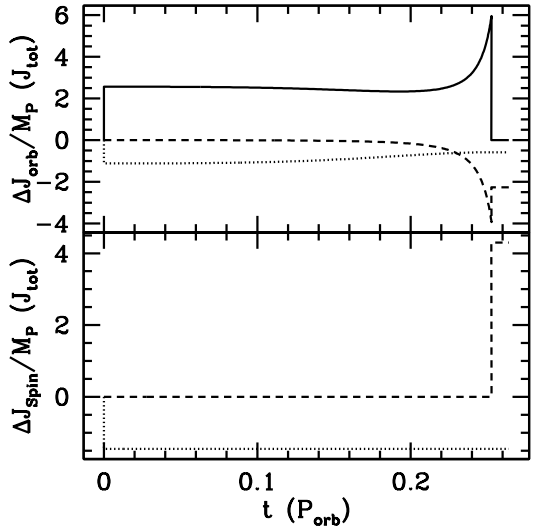}
\caption{Change in the angular momenta per unit ejected mass in units of 
the total angular momentum of the system as a function of time in units 
of the orbital period for a portion of a single ballistic orbit.  The 
shown system has a donor mass of $M_D=0.1\,M_\odot$, accretor mass of 
$M_A=0.45\,M_\odot$, and assumes the donor exactly fills its Roche lobe 
for a zero-temperature white dwarf mass-radius relation.  At top, we 
show the change in orbital angular momentum of each component.  At 
bottom we show the change in spin angular momentum of each component.  
The solid, dotted, and dashed lines show the changes in the angular 
momenta of the particle, donor, and accretor, respectively.}
\label{fig-orbit}
\end{figure}

In Figure~\ref{fig-orbit} we show the changes in the angular momenta per 
unit particle mass as a fraction of the total angular momentum of the 
system for a single ballistic trajectory as a function of time in units 
of the orbital period.  The system shown has $M_D=0.1\,M_\odot$ and 
$M_A=0.45\,M_\odot$.  As described above, the semi-major axis was chosen 
such that the donor star completely fills its Roche lobe for a 
zero-temperature mass-radius relation.  As the particle is ejected at 
time $t=0$ we see an instantaneous decrease in the orbital angular 
momentum of the donor star (dotted lines), with an accompanying decrease 
in its spin angular momentum.  The accretor's orbital and spin angular 
momenta (dashed lines) remain unchanged.  The orbital angular momentum 
of the particle (solid line) increases instantaneously.

Between ejection ($t = 0$) and accretion ($t\approx 0.25$), the orbital 
angular momentum of the particle, donor, and accretor vary.  During the 
first half of the particle transport ($t \lesssim 0.1$) there is very 
little change in any orbital angular momenta since the particle remains 
very close to the inner Lagrangian point of the donor, and hence the net 
force on the particle is small.  As the particle falls towards the 
accretor it more than doubles its orbital angular momentum, which is 
removed, primarily, from the accretor.  The orbital angular momentum of 
the donor increases slightly during this segment.

At accretion, the entirety of the particles orbital angular momentum is 
deposited into the accretor.  But, as can be seen in the Figure, not all 
of that momentum is deposited into the spin of the accretor.  For the 
system shown, the increase in the orbital angular momentum of the donor 
is approximately $1/2$ of the increase in its spin angular momentum.  

The proportion deposited into the spin versus the orbit depends directly 
on the angle between the normal to the surface at the point of impact 
and the instantaneous velocity of the particle at impact.  If the 
particle impacts exactly radially to the surface, the particle's angular 
momentum is deposited only into the orbital angular momentum of the 
accretor.  If the particle's velocity is exactly tangential to the 
surface at impact, then the particle's angular momentum is deposited 
entirely into the spin of the accretor. If the particle's velocity is at 
an angle neither tangential nor perpendicular to the surface, then the 
particle's angular momentum is distributed between the orbital and spin 
angular momenta as determined by equations~(\ref{eq-dj})-(\ref{eq-ac}).

During impact, the orbital and spin angular momenta of the donor remain 
unchanged.

In this example system, after a complete cycle of particle ejection from 
the donor, motion from donor to accretor, and accretion, the orbital 
angular momenta of the donor and accretor, and thus of the system as a 
whole, decreased. The spin angular momentum of the donor decreased, 
while the spin angular momentum of the accretor increased.  In 
section~\ref{sec-disc} we show the total changes in the spin and orbital 
angular momentum of each object, as well as the system as a whole, 
across broad parameter space for Roche lobe overflow double white dwarf 
binaries.

We note that any changes in the orbital angular momentum in this fashion 
are bound to instantaneously change the eccentricity of the system.  
While such an eccentricity is not relevant to the results of this paper, 
it is reasonable to assume that since the accretion is expected to be 
symmetric with orbital phase any induced eccentricity will be 
conservatively damped, resulting in a system with the same orbital 
angular momentum in a circular orbit.  We defer such eccentricity
considerations to future investigations.

\subsection{Applicability of the Ballistic Motion Assumptions}

Finally, we note that changes in the properties of the binary due to 
mass transfer (equations~[\ref{eq-dr}]--[\ref{eq-dj}]) are directly 
proportional to the mass of the ballistic particle.  Thus, for $M_P \ll 
M_D,M_A$, ejection of the ballistic particle should have little to no 
effect on the orbital changes of immediately subsequent ballistic 
ejections.  Provided the total amount of mass transferred remains small, 
the rate of change of the binary properties can be determined by 
multiplying equations~(\ref{eq-dr})--(\ref{eq-dj}) by the mass transfer 
rate.  Even though the mass transfer rate may be high 
during Roche lobe overflow \citep[$\lesssim 
10^{-5}\,M_\sun\,yr^{-1}$, ][]{2004MNRAS.350..113M}, the orbital periods 
are very short ($\lesssim$ hours).  This implies that amount of mass 
ejected each orbit should be of order $M_P/M_D \approx 10^{-9}$.  While 
the orbital parameters will not change appreciably during each orbit, 
they will slowly evolve on the timescale of years.  
\citet{2004MNRAS.350..113M} shows the evolution of the mass transfer 
rate in the case where the donor spin remains fixed (see 
\S\ref{sec-comp}).  The evolution of the mass transfer rate given 
the model presented here will be presented in a forthcoming paper 
\citep{Kremer2013}.

\section{Results}
\label{sec-angmom}

\subsection{Angular Momentum Changes Due to Direct Impact Accretion}
\label{sec-nonsync}

\begin{figure}
\plotone{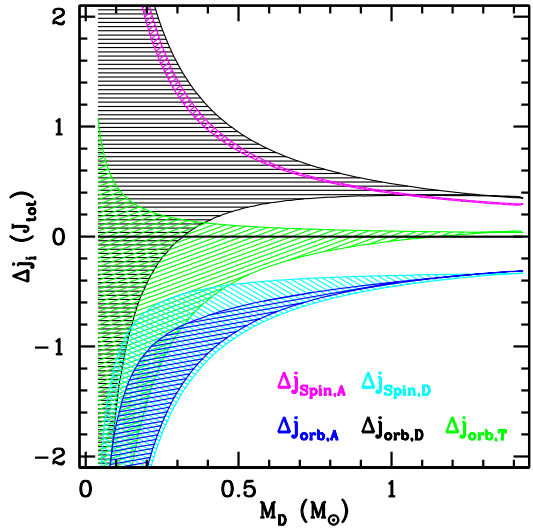}
\caption{The single-orbit change in the orbital angular momentum per 
unit transferred mass in units of the total angular momentum as a 
function of the donor mass for white dwarfs undergoing DI mass transfer.  
For each donor mass, the vertical extent of the shaded regions 
correspond to variations in the accretor mass that can be seen in 
Figure~\ref{fig-di}.  The line-shaded regions correspond to the changes 
in the angular momentum associated with a specific component of the 
system as follows: cyan and magenta are the change in the spin angular 
momentum per unit transferred mass for the donor and accretor, 
respectively; black and blue are the change in the orbital angular 
momentum per unit transferred mass for the donor and accretor, 
respectively; and green is the total change in the orbital angular 
momentum of the system per unit transferred mass.  The solid black line 
consistent with $\Delta j=0$ shows the change in the total angular 
momentum per unit transferred mass.}
\label{fig-jall}
\end{figure}

In Figure~\ref{fig-jall} we show the change in angular momentum per unit 
ejected mass in units of the total angular momentum following a single 
ballistic orbit as a function of the donor mass.  Only systems 
undergoing DI mass transfer as identified in Figure~\ref{fig-di} are 
shown.  The shaded regions of each color show the variation in systems 
as a function of accretor mass.  Shown in cyan and magenta are the 
changes in the spin angular momenta per unit transferred mass of the 
donor ($\Delta j_{\rm Spin, D}$) and accretor ($\Delta j_{\rm Spin, 
A}$), respectively; in black and blue are the changes in the orbital 
angular momenta per unit transferred mass of the donor ($\Delta j_{\rm 
orb, D}$) and accretor ($\Delta j_{\rm orb, A}$), respectively; and in 
green is the change in the total orbital angular momentum of the system 
($\Delta j_{\rm orb, T} = \Delta j_{\rm orb, A} + \Delta j_{\rm orb, 
A}$).  The black solid line shows the change in the total angular 
momentum of the system.  This line is consistent with zero to 
better than 1 part in $10^5$ throughout the parameter space and shows 
the accuracy of our angular momentum calculations.

A few characteristics of Figure~\ref{fig-jall} deserve more discussion.

First, we see that the change in the spin angular momentum of the 
accretor (magenta) is always positive.  This is as expected from the 
literature, where the ballistic particle impacts the surface of the 
accretor with a tangential velocity greater than the rotational velocity 
at the surface of the accretor.  Repeated accretion of this fashion will 
eventually create a rapidly rotating accretor.  We see that, as the 
donor mass decreases, the change in the spin angular momentum of the 
accretor increases significantly.  Furthermore, we note that this 
parameter has only a minor dependence on the mass of the accretor.

Next, we see that the change in the spin angular momentum of the donor 
(cyan) is always negative.  Since the particle is being ejected from the 
$L_1$ point with velocity components due to the rotational velocity of a 
uniformly rotating donor at that point, the particle carries away the 
specific spin angular momentum of the surface of the donor plus a small 
contribution due to the thermal ejection velocity.

Similarly, the change in the orbital angular momentum of the accretor 
(blue) is always negative.  The variation is due to the position of the 
particle impact, as well as the direction of the particle's impact 
velocity and the other binary properties.  Additionally, there can be a 
significant change in the orbital angular momentum of the accretor 
during the particle transport, as can be seen in Figure~\ref{fig-orbit}.

We see that the change in the orbital angular momentum of the donor 
(black) can be either positive or negative, depending on both the mass 
of the donor and of the accretor.  For a mass ratio $M_D/M_A > 0.27$, 
$\Delta j_{\rm orb, D}$ is always positive, though for systems $M_D/M_A 
< 0.27$ $\Delta j_{\rm orb, D}$ may either increase or decrease.  As 
with the change in the spin angular momentum of the donor, this will 
vary based on the velocity and angle of the particle ejection.  
Additionally, the orbital angular momentum of the donor may change 
during the particle motion, as can be seen in Figure~\ref{fig-orbit}.

The change in the total orbital angular momentum of the system (green) 
can be either positive or negative, depending on both the mass of the 
donor and of the accretor.  For a mass ratio $M_D/M_A > 0.85$, $\Delta 
j_{\rm orb, T} > 0$, though for systems $M_D/M_A < 0.85$ $\Delta j_{\rm 
orb, T}$ may either increase or decrease depending on the component 
masses.  We discuss the implications of $\Delta j_{\rm orb, T} > 0$ in 
\S\ref{sec-disc}.

We emphasize that the approaches to this problem previously presented in 
the literature (i) do not include any change in the orbital angular 
momentum of either the donor or accretor due to the ejection or 
accretion of the particle; and (ii) change the orbital angular momentum 
of the system during particle motion by adjusting the semi-major axis 
instead of accounting for the changes in orbital angular momentum of 
either/both component stars.  The results presented here do include such 
changes in the position and orbital and rotational velocities of both 
objects during ejection, particle motion, and accretion, which in turn 
create the observed changes in the angular momenta.  We defer 
investigations of the eccentricity induced during this process to a 
future study.

We note that, as discussed in the previous section, provided the mass of 
the ejected particle is sufficiently small, such changes may be scaled 
by the desired mass transfer rate to obtain the rate of change of the 
desired quantity as a function of time.  Furthermore, it is imperative 
to realize that any changes in the orbital or spin angular momenta of 
either component will affect any subsequent ballistic ejection, orbit, 
and accretion.  Thus, the changes shown above are {\it not} a steady 
state solution but merely represent the initial single-orbit change for 
the given set of initial parameters.  For an investigation of the 
long-term orbital changes due to the process described here, see 
\citet{Kremer2013}

\subsection{Angular Momentum Changes with Strong Tidal 
Coupling}
\label{sec-sync}

\begin{figure}
\plotone{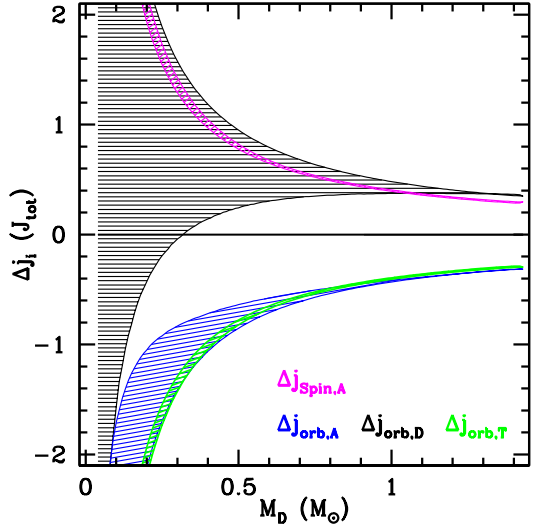}
\caption{As Figure~\ref{fig-jall}, but assuming the donor 
rotational velocity remains synchronized with the orbital velocity.}
\label{fig-jall-sync}
\end{figure}

It is commonly assumed (\citet{2004MNRAS.350..113M}; but see also 
\citet{2007ApJ...655.1010G}) that the rotational velocity of the donor 
will remain synchronized with the orbital velocity.  This is generally 
expected since Roche lobe overflow requires extreme tidal distortions in 
the donor star which act to synchronize its rotational velocity on short 
timescales.  The accretor is generally not affected as strongly by tidal 
synchronization since, being more massive, it is not expected to be 
significantly tidally distorted.  Some tidal coupling between the 
accretor and the orbit is generally assumed, though it is not expected 
to be strong enough to keep the accretor from becoming a rapid rotator.

To simulate a strong tidal coupling between the donor and the orbit, we 
show in Figure~\ref{fig-jall-sync} the same results as 
Figure~\ref{fig-jall} except we assume the donor's rotational velocity 
remains synchronized to the orbital velocity.  Hence, we assume the 
angular momentum that would be removed from the spin of the donor is 
immediately returned from the orbit.  In Figure~\ref{fig-jall-sync}, 
$\Delta j_{\rm orb, T} = \Delta j_{\rm orb, D} + \Delta j_{\rm orb, A} + 
\Delta j_{\rm Spin, D}$.  As expected, the total change in the orbital 
angular momentum in this case is simply the negative of the change in 
the spin angular momentum of the donor.  The changes in the orbital 
angular momenta of both components remains the same as 
Figure~\ref{fig-jall}.

\subsection{Comparison to previous results}
\label{sec-comp}

\begin{figure}
\plotone{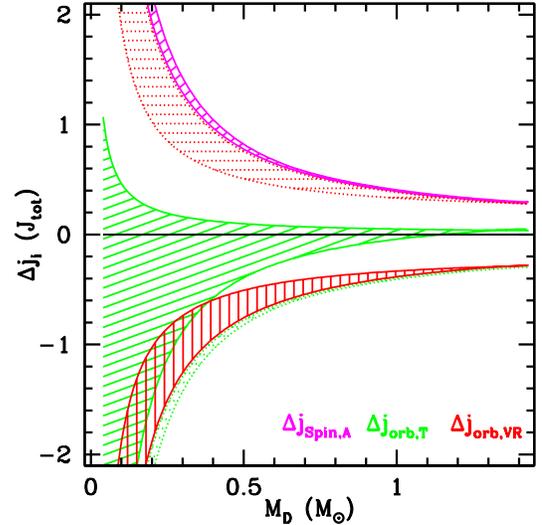}
\caption{As Figure~\ref{fig-jall}, but adding comparisons to the 
standard prescription from the literature.  The green line-shaded area 
corresponds to the change in the total orbital angular momentum per unit 
transferred mass for the approach presented in \S\ref{sec-nonsync}.  The 
green dot-shaded area corresponds to the change in the total orbital 
angular momentum per unit transferred mass for the approach presented in 
\S\ref{sec-sync}.  The magenta line-shaded area corresponds to the 
change in the spin angular momentum of the accretor per unit transferred 
mass for the approach presented in \S\ref{sec-nonsync}.  The red 
line-shaded region corresponds to the change in orbital angular momentum 
per unit transferred mass that would arise following the prescription of 
\citet{1988ApJ...332..193V}, while the red dot-shaded region corresponds 
to the change in the spin angular momentum of the accretor per unit 
transferred mass under the same assumptions. See text for details.}
\label{fig-jall-vr}
\end{figure}

In Figure~\ref{fig-jall-vr} we compare the change in the total orbital 
and accretor spin angular momenta per unit transferred mass for the 
non-synchronous and synchronous models presented earlier, as well as the 
change in the orbital angular momentum per unit transferred mass as 
presented in the literature.

The green line-shaded area shows the change in the total orbital angular 
momentum per unit transferred mass where we allow the spin angular 
momentum of the donor to change self-consistently, as shown in 
Figure~\ref{fig-jall}.  The green dot-shaded region shows the total 
orbital angular momentum per unit transferred mass where the rotation 
rate of the donor is fixed, as shown in Figure~\ref{fig-jall-sync}.  The 
magenta line-shaded region is the change in the spin angular momentum of 
the accretor per unit transferred mass.  The red line-shaded region 
shows the change in the angular momentum of the orbit per unit 
transferred mass used in \citet{2004MNRAS.350..113M} and 
\citet{2007ApJ...655.1010G}. The red dot-shaded region shows the spin 
angular momentum per unit transferred mass added to the accretor as used 
in the previous works and is simply the negative of the red dash-shaded 
region.

Following \citet{2004MNRAS.350..113M} we calculate the red dash-shaded 
region employing the method of \citet{1988ApJ...332..193V}.  They begin 
by writing the orbital angular momentum of the transferred mass in terms 
of the circular orbit with the same specific angular momentum.  They 
then use their equation~(13) as their fitting formula for the radius of 
this orbit.  The fitting formula is scaled to the semi-major axis of the 
orbit which we calculate by combining the fitting formula for the Roche 
lobe radius given by \citet{1983ApJ...268..368E} with the fitting 
formula for the radius of zero-temperature white dwarfs of Eggleton 
(1986) \citep[as quoted by][]{1988ApJ...332..193V}, assuming the donor 
completely fills its Roche lobe.  The fitting formula used, however, 
does not represent the angular momentum of the particle at impact.  
Instead, it is a fit to the angle-averaged radius of an ejected 
particle throughout a ballistic orbit as given by \citet[][$\varpi_d$; 
table~2]{1975ApJ...198..383L}.

We see that, for most of the parameter space, the change in the total 
orbital angular momentum of the system is more positive than when 
applying the prescription of \citet{1988ApJ...332..193V} (see 
Figure~\ref{fig-jall-vr}).  Thus, it 
appears that mass transfer using these ballistic assumptions may have 
less of a destabilizing affect that previously expected.  Furthermore, 
since there are areas of the parameter space where $\Delta j_{\rm orb, 
T}>0$, mass transfer may {\it increase} the stability of the orbit.

\begin{figure}
\plotone{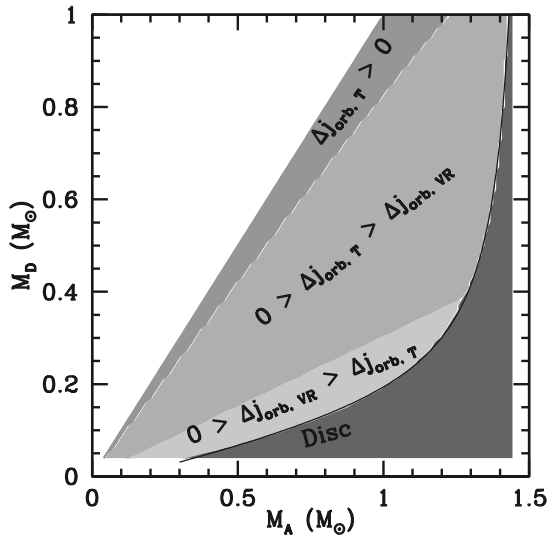}
\caption{As Figure~\ref{fig-di}, but we further divide the DI 
parameter space into regimes based on the net change in 
orbital angular momentum following a ballistic ejection, particle 
motion, and accretion.  The darkest grey region undergoes DF.  In the DI 
regime, from darkest to lightest, the regions represent: (1) where the 
total orbital angular momentum increases, (2) where the total orbital 
angular momentum decreases by an amount less than that assumed by 
previous studies, (3) where the total orbital angular momentum 
decreases by a greater amount than that assumed by previous studies.  
The thin black line shows the fit for the transition between DI and DF 
given by equation~\ref{eq-fit}.}
\label{fig-di-j}
\end{figure}

In Figure~\ref{fig-di-j} we show the changes in the orbital angular 
momentum per unit transferred mass in the parameter space of 
Figure~\ref{fig-di}.  As in Figure~\ref{fig-di}, the dark grey region in 
the lower right corresponds to systems which undergo DF.  The remainder 
of the parameter space are systems which undergo DI, which we further 
subdivide, from darkest to lightest, as follows: (1) systems where the 
total orbital angular momentum increases; (2) systems where the total 
orbital angular momentum decreases, but at a slower rate than previously 
assumed in the literature; (3) systems where the total orbital angular 
momentum decreases more rapidly than previously assumed in the 
literature.  The dividing lines between each region are nearly, but not 
quite, linear.  The dividing line between regions (1) and (2) has a 
slope of approximately $M_D/M_A \approx 0.825$, while the dividing line 
between regions (2) and (3) has a slope of approximately $M_D/M_A 
\approx 0.30$.  Thus, the change in the total orbital angular momentum 
appears to become more positive with increasing donor mass.

Curiously, this stability is directly opposite that seen in previous 
investigations.  The analysis of \citet[][e.g., 
Figure~1]{2004MNRAS.350..113M} shows that systems with a low donor mass 
are more likely to remain stable over long periods of time, while the 
majority of the parameter space (depending on the strength of the tidal 
coupling) is expected to be unstable.  Here, by including the additional 
mass transfer effects presented here in a self-consistent way, DWD DI 
mass transfer may induce a stabilizing effect over a larger area of the
parameter space. Indeed, as seen in Figure~\ref{fig-di-j}, the region 
furthest from the stability region of \citet{2004MNRAS.350..113M} 
receives the largest stabilizing effect, where the orbital angular 
momentum of the system is {\it increased} due to DI mass transfer.  As 
can be seen from Figure~\ref{fig-jall}, this increased orbital angular 
momentum arises at the expense of the spin angular momentum of the 
donor and the ejection velocity of the ejected mass, which provides a 
small impulse to the center of mass of the donor, pushing it to a larger 
orbit and increasing its orbital angular momentum.  The net 
effect of DI mass transfer, when tracking the orbital angular momentum 
of the system in detail, is a system more stable than that expected 
following the assumptions in the current literature.

\section{Discussion}
\label{sec-disc}

We have shown that, if one takes into account the changes in both the 
spin and orbital angular momenta of the component stars during mass 
ejection/accretion, as well as numerically calculating the ballistic 
model for each mass transfer scenario, the resultant transfer of angular 
momentum can be significantly different than the result currently 
described in the literature.  In many cases the orbital angular momentum 
lost from the orbit can be significantly less than the standard 
assumption, making this process less destabilizing than expected.  This 
may allow for more DWD to survive the semi-detached state and become AM 
CVn, instead of merging to create Type Ia or Type .Ia supernovae.  This 
may have significant implications on population synthesis models of 
these objects.

In a few cases, we have shown that mass transfer may {\it increase} the 
orbital angular momentum of the orbit, thereby providing a stabilizing 
effect on the orbit.  We caution again that the results presented here 
are for a single mass transfer event over a single binary orbit, and 
thus do not represent a steady state solution.  Nevertheless, any 
stabilizing effect increases the chances of a long-lived semi-detached 
DWD, lending credence to the creation of AM CVn through the DWD 
scenario.

In order to analyze changes in the long-term stability of these systems, 
we will incorporate our ballistic calculations into a long term 
numerical integration, taking into account tides and gravitational 
radiation.  By updating the ballistic trajectories and angular momentum 
changes at every timestep, we will be able to self-consistently 
calculate the long-term evolution of the system conserving the total 
angular momentum from ejection to accretion.  Results from these 
calculations will be presented in a forthcoming paper 
\citep{Kremer2013}.

\acknowledgements
The authors acknowledge many useful conversations and debates which 
helped to shape the work into its present form, particularly those with 
Tom Marsh, Bart Willems, Paul Groot, Danny Steeghs, Gijs Nelemans, and 
Juhan Frank.  This project was supported in part by an internal faculty 
research grant from the University of Scranton to JS, a Simons 
Fellowship in Theoretical Physics to VK. VK is also grateful for the 
hospitality of the Aspen Center for Physics where part of this work was 
completed.

\bibliography{MT}

\end{document}